# Enhancing Fault Detection in CO2 Refrigeration Systems: Optimal Sensor Selection and Robustness Analysis Using Tree-Based Machine Learning

Masoud Kishani Farahani[a], Morteza Kolivandi[a],
Abbas Rajabi Ghahnavieh[a,*], Mohammad Talaei[a]

[a] Department of Energy Engineering, Sharif University of Technology, Tehran, Iran, E-mail: rajabi@sharif.edu

## Abstract

This study investigates the reliability and robustness of data-driven Fault Detection and Diagnosis (FDD) models for CO2 refrigeration systems (CO2-RS) in supermarkets, focusing on optimal sensor selection and resilience against sensor noise. Employing tree-based machine learning algorithms—Random Forest (RF), XGBoost, CatBoost, and LightGBM—we developed FDD models to classify six common faults in a laboratory-scale CO2-RS. The Recursive Feature Addition (RFA) approach identified optimal sensor sets, achieving a 99% F1-score with minimal sensors: four for RF, seven for XGBoost, five for CatBoost, and five for LightGBM. Condenser-side sensors consistently ranked as critical for fault detection. Robustness was assessed by injecting Additive White Gaussian Noise (AWGN) at a signal-to-noise ratio (SNR) of 3 dB into the most important sensor, with XGBoost demonstrating superior resilience at 85.24%, followed by CatBoost (57.07%), LightGBM (49.1%), and RF (49.46%). Sensitivity analysis across high-SNR (10 dB), low-SNR (0 dB), and sensor failure scenarios revealed XGBoost's robustness peaking at 90.23% and retaining 79% under failure, contrasting with sharper declines in other models. These findings highlight a trade-off between sensor count, cost, and reliability, with larger ensembles enhancing noise resilience. This work bridges gaps in FDD literature by integrating sensor optimization with comprehensive robustness analysis, offering a practical framework for improving energy efficiency and fault management in CO2-RS. Future efforts could explore adaptive SNR thresholds and redundant sensor configurations to enhance real-world applicability.

## Nomenclature

| | |
|---|---|
| $CO_2$-RS | $CO_2$ Refrigeration Systems |
| FDD | Fault Detection and Diagnostics |
| ML | Machine Learning |
| RFA | Recursive Feature Addition |
| SNR | Signal-to-Noise Ratio |
| LT | Low-Temperature |
| MT | Medium-Temperature |
| RF | Random Forest |
| MDI | Mean Decrease in Impurity |
| AWGN | Additive White Gaussian Noise |
| $T_{sup1}$, $T_{sup2}$ | LT/MT evaporator supply air temperature |
| $T_C$ | Condenser outlet temperature |
| $T_{ret1}$ | MT evaporator return air temperature |
| $T_{suc3}$, $T_{suc5}$ | MT/LT 3$^{rd}$/2$^{nd}$ compressor suction temperature |
| $T_{suc7}$ | LT compressor rack inlet temperature |
| $T_{dis7}$ | LT compressor rack outlet temperature |
| $T_{FI}$, $T_{FO}$ | Condenser Fan Inlet/Outlet air temperature |

## 1. Introduction

The 2024 global status report for buildings and construction highlights that the building sector is responsible for around 21% of global greenhouse gas emissions, with buildings in Hong Kong consuming 90% of the city's electricity and accounting for over 60% of its carbon emissions [1,2]. Supermarkets stand out as high-energy consumers in the building sector due to their heavy reliance on refrigeration systems [3]. Supermarket owners have turned to $CO_2$ refrigeration systems ($CO_2$-RS) with the aim of reducing energy consumption and greenhouse gas emissions, as they have a low environmental impact and consume less energy [4]. However, $CO_2$-RS are particularly susceptible to operational faults, which can undermine their advantages. Studies have shown that 15% to 30% of total energy consumption in commercial buildings may be wasted due to insufficient maintenance and a lack of monitoring for operational faults [5]. Therefore, it is crucial to identify solutions that detect operational faults and address malfunctioning issues. Lee et al. demonstrated that, on average, 14.1% of the energy consumed by equipment in buildings could be saved if the equipment were operated correctly [6].

Fault Detection and Diagnosis (FDD) plays a pivotal role in identifying and addressing operational faults of $CO_2$-RS, thereby reducing energy and maintenance expenses and contributing to global warming mitigation. Table 1 summarizes key research studies on FDD in $CO_2$-RS.

Table 1. Summary of research studies on FDD in $CO_2$-RS.

| **Authors** | **Focus** | **Methodology/Approach** | **Key Findings** | **References** |
|---|---|---|---|---|
| Li et al. (2022) | Detection and diagnosis of common faults | Gray-box approach | Achieved over 90% accuracy in fault detection. | [7] |
| Sun et al. (2021) | Development of a characteristic matrix for fault diagnosis | Fault analysis | Identified four main faults to aid in FDD model development. | [8] |
| Sun et al. (2022) | Implementation of automated FDD system | IoT solution platform | Deployed an automated FDD system via cloud-based service. | [9] |
| Farahani et al. (2025) | Development of data-driven FDD models | Tree-based ML algorithms | Classified six common faults with up to 99.48% accuracy. | [10] |

Zhang et al. [11] underscored two research gaps in FDD literature: research on data-driven FDD models has now centred more on ML algorithms than on selecting an optimal sensor set for them, and the effect of sensor noise on FDD models is not well studied. Generally, sensors play a pivotal role in developing data-driven FDD models by providing information about various aspects of the system's behaviour through monitoring. On the other hand, selecting the optimal subset of sensors is a crucial component in the process of developing accurate and efficient FDD models [12]. This component is a collection of sensors that offers the most essential data for data-driven FDD models, enabling them to effectively detect and diagnose faults while utilizing the least number of sensors. The process of identifying an optimal sensor set is analogous to feature selection practices in other ML disciplines. In this context, Recursive Feature Addition (RFA) is an iterative feature selection approach for finding an optimal feature set that leads to improved FDD models performance. This approach is a ”forward selection”, where features are added one by one to the classifier based on their importance until the predefined performance criteria are met [13]. Lu et al. validated RFA's effectiveness in selecting optimal sensor sets using feature importance rankings for a chiller [14].

Although feature selection methods, such as the RFA approach in FDD model development, assist in selecting an optimal sensor set and reducing the initial costs of the models, very few studies have addressed the reliability of this optimal sensor set. While an optimal sensor set can indeed reduce the implementation costs of FDD models in practical applications, it does not guarantee reliability and robustness against environmental noise. As Zhang et al. [11] pointed out in a comprehensive systematic review of existing research encompassing more than 100 articles related to FDD model development and construction, the reliability and robustness of FDD models developed with an optimal sensor set represent significant research gaps in the FDD literature. They noted that sensor noise is an important concern in FDD, and they found only two papers that discuss it [15,16]. Bonvini et al. [16] presented a robust and capable FDD algorithm that has been tested against various levels of noise in sensor data to evaluate the algorithm's robustness with respect to noisy and erroneous data. However, no study has simultaneously investigated the robustness of FDD models developed with an optimized sensor set against sensor noise.

Therefore, one of the main important aspects of developing FDD models is finding out how sensor noise affects their performance. Sensor noise is the result of random variations in sensor output and can be caused by various factors, such as environmental disturbances, device imperfections, or signal interference. This can lead to false alarms, missed detections, and reduce the accuracy and reliability of FDD models [17, 18, 19]. These disturbances caused by noise are usually measured through the signal-to-noise ratio (SNR). To analyse the effect of sensor noise on the performance of FDD models, a concept called robustness analysis is used. Robustness analysis is a concept in FDD that refers to the ability of FDD models to maintain their performance versus sensor noise under specified SNR levels [20].

The primary objective of this study is to investigate the reliability of optimal sensor set and robustness analysis of FDD models against sensor noise. The intended outcomes aspire to foster increased FDD model adoption within the supermarket sector, concurrently bridging prevailing research gaps. The main contributions of this study are summarized below:

1) **Optimal sensor selection:** A novel approach to selecting an optimal sensor set is proposed to identify essential sensors for precise fault detection in $CO_2$-RS. This contribution minimizes sensor count while maintaining the FDD model's performance and robustness.
2) **Robustness analysis against sensor noise:** The study presents a comprehensive robustness analysis of FDD models against sensor noise. The findings provide insights into quantifying the impact of sensor noise on FDD models by injecting controlled noise. Controlled noise injection under a specified SNR level contributes to a deeper understanding of the FDD model's robustness.

The data-driven method tree-based ML algorithms are applied for building FDD models in this work. For selecting an optimal sensor set, we apply the RFA approach and the feature importance ranking technique based on tree-based ML algorithms. The process of robustness analysis of FDD models against sensor noise and the noise injection under a specified SNR level is described in detail in section 3. The remainder of this study is structured as follows. First, a commercial $CO_2$ refrigeration system was selected as a case study, as well as faults and sensors that were installed on it are introduced in section 2. Section 3, Methodology, encompasses the comprehensive steps taken for selecting an optimal sensor set. In section 4, (Result and Discussion) the performance and robustness of the built FDD models are compared and evaluated in the presence and absence of sensor noise.

## 2. Case Study

A laboratory-scale $CO_2$-RS was selected as the case study to validate the proposed methodology. As shown in Figure 1 [21], the system operates in a two-stage configuration, delivering both Low-Temperature (LT) and Medium-Temperature (MT) cooling. The system has a total cooling capacity of 43 kW, with LT and MT loads simulated using plate heat exchangers, electric heaters, and glycol loops to mimic heat sources. Six common faults identified from Reference [21] were tested in an experimental setup, and datasets were created to analyze their impact on system performance. The selected faults include an open LT display case door, ice accumulation on LT evaporator coil, LT evaporator expansion valve failure, MT evaporator fan motor failure, condenser air path blockage, and MT evaporator air path blockage. For each fault condition, data were recorded under both faulty and normal operating scenarios, ensuring consistent conditions throughout the experiment.

The experimental setup included 24 temperature sensors, 7 pressure sensors, and 3 refrigerant mass flow rate sensors installed at critical points, such as compressor inlets and outlets, display cases, expansion valves, and the condenser. Additionally, 6 power consumption sensors were installed on the compressors and condenser to monitor energy usage. These sensors provided continuous monitoring and recorded real-time data, which was stored in CSV files. Sensor locations are detailed in Figure 1, while Table A1 in the appendix lists the sensor types and abbreviations. The collected data were merged into a single dataset to develop data-driven FDD models.

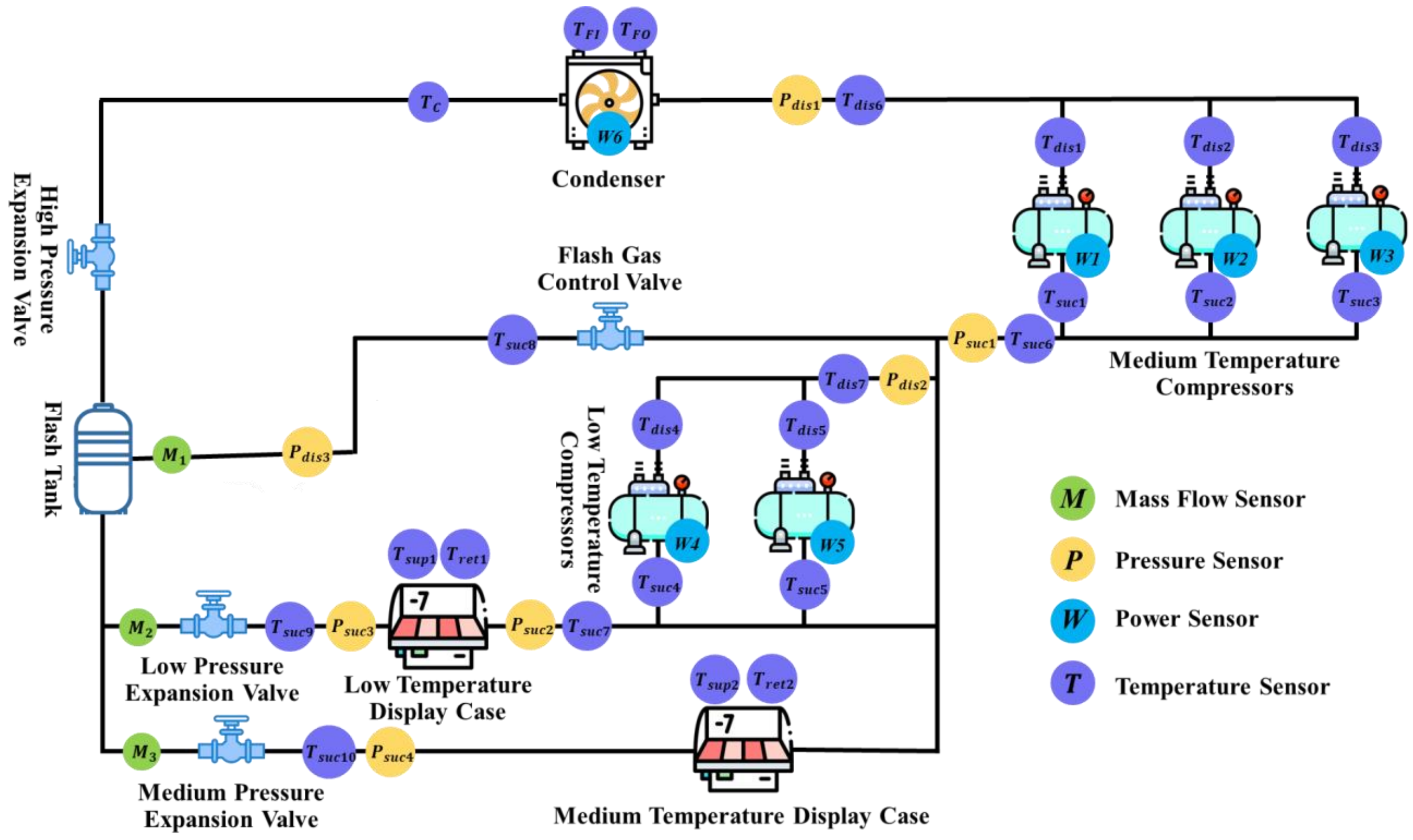

Figure 1. $CO_2$ refrigeration system diagram.

## 3. Methodology

### 3.1. Tree-Based Machine Learning Algorithms

Tree-based ensemble learning techniques leverage the hierarchical structure of decision trees to model complex relationships in data, offering robust solutions for classification and regression tasks [22]. This section details two primary approaches—Bagging and Boosting—and their mechanisms for ranking feature importance, which are integral to developing FDD models for fault classification using sensor data and for selecting an optimal set of sensors. Figure 2 illustrates the structure of tree-based ML algorithms.

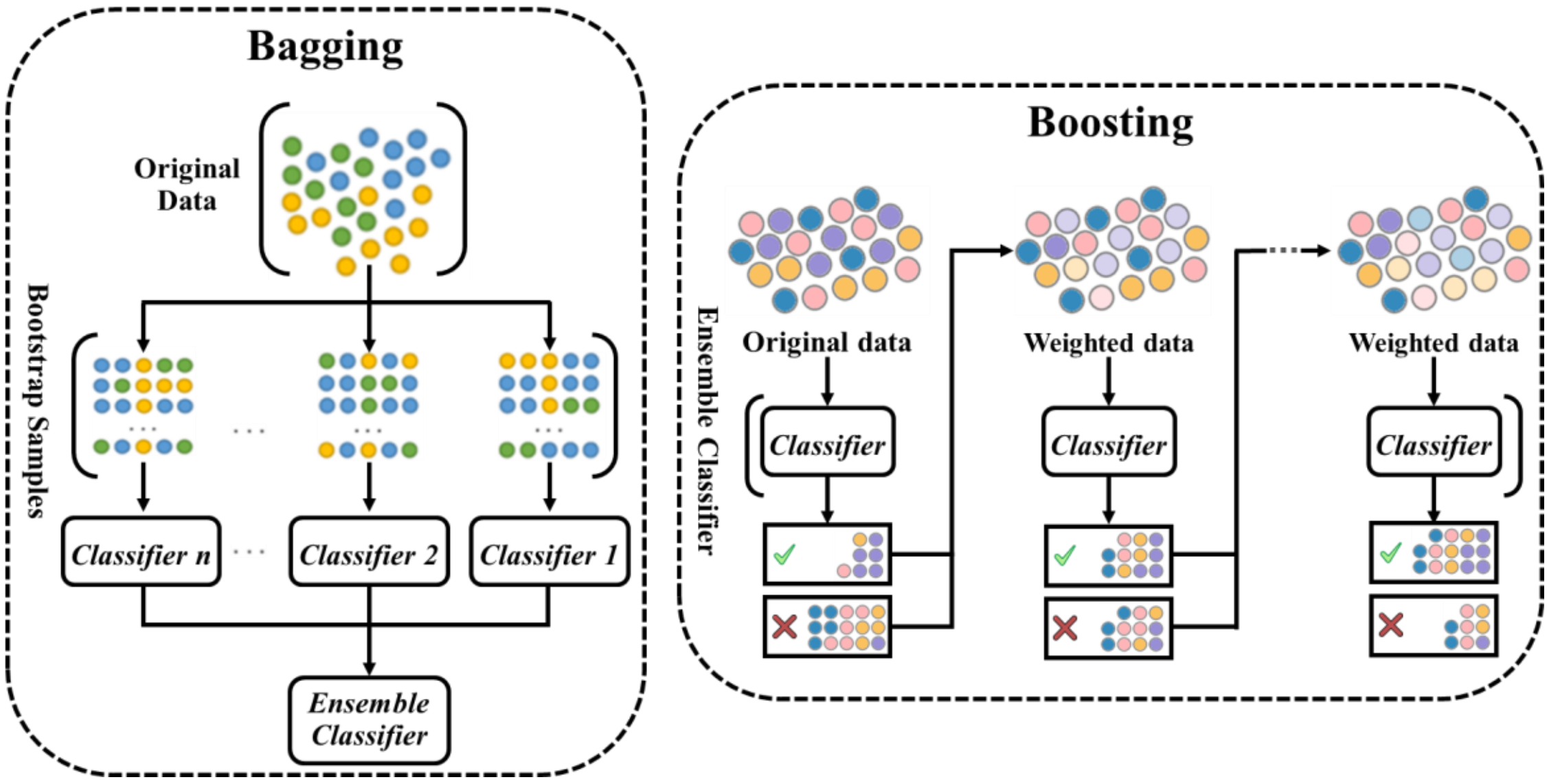

Figure 2. The structure of tree-based ML algorithms.

### 3.1.1 Bagging

Bagging, or Bootstrap Aggregating, enhances model performance by reducing variance through the aggregation of multiple decision tree predictions [23]. The technique generates diverse training subsets via bootstrapping—random sampling with replacement—and trains a separate decision tree on each subset independently. Predictions from these trees are combined, typically by averaging for regression or majority voting for classification, to produce a final output. This process is exemplified by the Random Forest (RF) algorithm, which introduces additional randomness by selecting a subset of features at each node to determine the optimal split. The aggregated prediction for RF is expressed as shown in Eq. (1).

$$\hat{y} = \frac{1}{N}\sum_{i=1}^{N} \hat{y}_i \qquad (1)$$

Where:
$N$ is the total number of trees in the forest.
$\hat{y}_i$ is the prediction from the $i$-th decision tree.
By averaging diverse tree outputs, Bagging mitigates overfitting and improves generalization across unseen data.

### 3.1.2 Boosting

In contrast, Boosting builds an ensemble sequentially, focusing on correcting errors from prior iterations to minimize bias [24]. Each decision tree, or weak learner, is trained on a weighted dataset where misclassified instances from previous trees receive increased emphasis. The final prediction aggregates contributions from all learners, scaled by a learning rate $\eta$, as shown in Eq. (2).

$$\hat{y} = \sum_{i=1}^{N} \eta \cdot \hat{y}_i \tag{2}$$

Where N is the number of learners, $\hat{y}_i$ is the prediction from the $i$-th learner.

Popular implementations, such as XGBoost, CatBoost, and LightGBM, optimize this process by employing gradient-based loss minimization, efficient handling of categorical features, or histogram-based split finding, respectively. Boosting's iterative refinement enhances accuracy, particularly for challenging data patterns.

### 3.1.3 Feature Importance Ranking

Both Bagging and Boosting provide methods to rank feature importance, offering insights into the predictive relevance of input variables [25]. In the case of Bagging, feature importance is typically assessed using the Mean Decrease in Impurity (MDI) method, particularly in RF models. This approach evaluates the contribution of each feature at every split by measuring the reduction in impurity, which is commonly quantified using Gini Impurity. The Gini Impurity is defined in Eq. (3).

$$I_G = 1 - \sum_{i=1}^{C} P_i^2 \tag{3}$$

Where:

$C$ is the number of classes.

$P_i$ is the proportion of samples belonging to class $i$ in the node.

The total importance of feature $X_j$ across all trees is computed in Eq. (4).

$$FI_{MDI}(X_j) = \frac{1}{N} \sum_{t=1}^{N} \sum_{s \in S_j} \Delta I_G(S, t) \tag{4}$$

Where:

$N$ is the total number of trees.

$S_j$ represents the set of all splits where feature $X_j$ is used.
$\Delta I_G(S,t)$ is the decrease in impurity at split $s$ in tree $t$.

This method provides insight into which features are most influential in determining the outcome and enable effective feature selection and ranking.

In the case of Boosting, feature importance is typically calculated via Gain-Based Importance focusing on the cumulative gain in loss reduction from splits on a feature. For a feature $X_j$, importance is computed in Eq. (5):

$$FI_{Gain}(X_j) = \frac{\sum_{t=1}^{N} \sum_{s \in S_j} Gain\ (s,t)}{\sum_{t=1}^{N} \sum_{s \in S} Gain\ (s,t)} \quad (5)$$

Where:
$Gain\ (s,t)$ is the improvement in the objective function (loss reduction) from splitting on feature $X_j$ at node s in tree t.
$S_j$ represents the set of all splits where feature $X_j$ is used.
$S$ is the set of all splits across all features.

These techniques—Bagging for variance reduction and Boosting for bias minimization—combined with their feature-ranking capabilities, enable the construction of robust, interpretable models suited to complex data-driven tasks.

### 3.2. Construction and Evaluation of Data-Driven FDD Models

In this study, we develop data-driven FDD models using tree-based ML algorithms. These models function as multi-class classification systems, trained on datasets to identify various fault types through supervised learning. We employ four tree-based algorithms— RF, XGBoost, CatBoost, and LightGBM—chosen for their proficiency in distinguishing fault-free states from multiple fault conditions in a multi-class framework. The construction process starts by splitting the dataset into training and testing subsets. Each algorithm is implemented as a classifier, and once the classifier is trained, the resulting FDD model becomes capable of classifying incoming data as either fault-free or indicative of a specific fault type.

After construction, the FDD models are evaluated using the testing subset to evaluate their diagnostic reliability. Performance evaluation is crucial in data-driven FDD systems to confirm that the models generalize effectively to unseen data and perform well in practical fault diagnosis scenarios. For this, we rely on the macro-average F1-score as the primary metric. This metric, derived from the *classification_report* function in *sklearn* library, is well-suited for FDD applications as it balances precision and recall across all classes, providing a comprehensive measure of performance, particularly when class distributions are uneven. The macro-average F1-score is calculated using Eqs. (6) and (7).

$$Macro\ F1\ score = \frac{1}{N}\sum_{i=1}^{N} F1\ score_i \quad (6)$$

$$F1\ score_i = \frac{2 \times \frac{TP_i}{TP_i + FP_i} \times \frac{TP_i}{TP_i + FN_i}}{\frac{TP_i}{TP_i + FP_i} + \frac{TP_i}{TP_i + FN_i}} \quad (7)$$

where $TP_i$ (true positives) are correctly classified samples of class $i$, $FP_i$ (false positives) are samples incorrectly classified as class $i$, and $FN_i$ (false negatives) are samples of class $i$ misclassified as another class.

The macro-average approach calculates the F1-score for each fault class individually and then averages them, giving equal weight to all classes irrespective of their sample size. This ensures a fair assessment of the model's ability to detect and diagnose all fault types, making it an ideal choice for FDD models where both missing faults (low recall) and falsely identifying fault-free states as faulty (low precision) are critical concerns. The evaluation is conducted on the testing set, verifying the model's performance on unseen data.

### 3.3. Optimal Sensor Set Selection Method

The RFA approach is a systematic method for feature selection that incrementally adds features, beginning with the most important, to construct an optimal feature subset for FDD. Combined with a feature importance ranking technique, which evaluates each feature's contribution to the classification process, RFA ensures that only the most relevant features are included in the model. This approach not only enhances the model's diagnostic performance but also reduces computational complexity and improves interpretability by focusing on the key contributors to fault classification.

To identify the optimal sensor set, the following steps are carried out using the RFA approach:

1. Incrementally adding sensors, one by one, to a new dataset based on their importance ranking from the sorted dataset.
2. Splitting the new dataset into training and testing sets.
3. Training classifiers on the training set using the added sensors to construct a data-driven FDD model.
4. Evaluating the constructed FDD model's performance on the testing set and storing the results.
5. Comparing the obtained FDD model's performance at each step of sensor addition against a predefined threshold or stopping criterion (e.g., F1_Score = 99%).

The process continues until the predefined threshold is achieved. The flowchart outlining this process is illustrated in Figure 3. For example, if the model's performance exceeds the threshold after adding a new feature (compared to the previous model with fewer features), the added feature is considered important and retained. The stopping criterion could be defined by the user as a specific performance level. If the criterion is not met, the new dataset is updated with the next sensor, and the addition process continues. Ultimately, the most important and influential sensors are selected to form the optimal sensor subset.

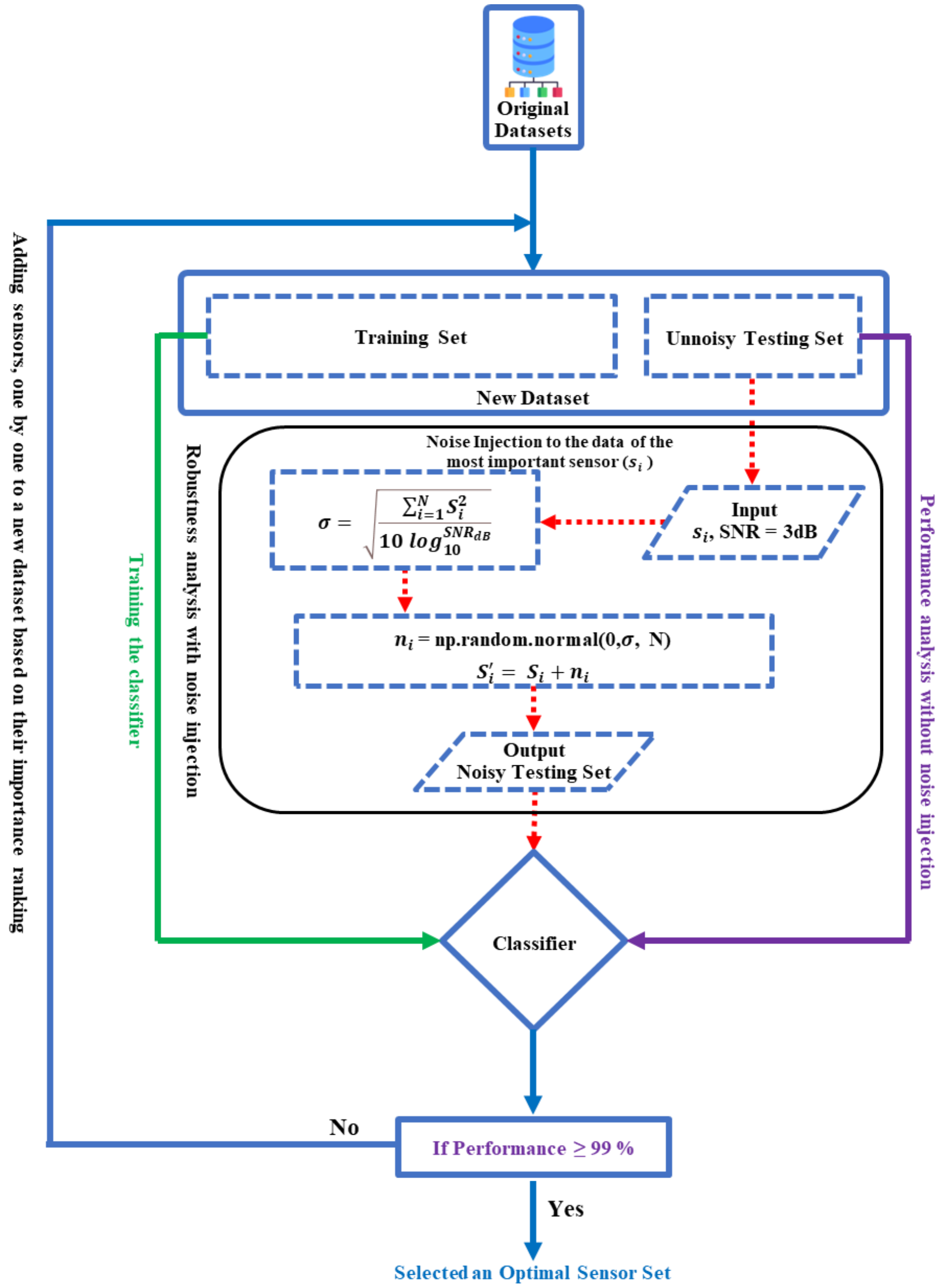

Figure 3. The proposed methodology for selecting an optimal sensor set and robustness analysis.

### 3.4. Robustness Analysis

Robustness analysis is crucial to evaluate the impact of sensor noise on FDD model performance. Sensor noise can have a significant impact on the performance of FDD models by introducing noise in the sensor data used to detect and diagnose faults in a system. Data-driven FDD models learn based on input information from sensor data. SNR can be used to evaluate the quality of sensor data to determine whether the data is informative enough so FDD models can detect faults reliably. The higher SNR, the better the quality of the sensor data and the lower the impact of noise on the data. SNR is quantified in decibels (dB) and is defined as Eq. (8).

$$SNR_{dB} = 10log_{10}^{\frac{P_{signal}}{P_{noise}}} \qquad (8)$$

where $P_{signal}$ and $P_{noise}$ are the power of the signal and the noise, respectively.

However, no minimum range for SNR has been reported in studies for data acquisition systems such as sensors. In this regard, to analyze the robustness of the built models against sensor noise, we assume a minimum SNR equal to 3dB. In this SNR, the signal power is approximately 2 times the noise power based on Eq. (8). Assuming a minimum SNR of 3dB is a reasonable approach and starting point for analyzing the robustness of FDD models against sensor noise. With this SNR, we can be sure that the FDD models are able to perform well even in noisy environments. A lower SNR would make it more difficult for the models to correctly identify the signal, and a higher SNR would be unnecessary. However, the minimum SNR for FDD models varies depending on the specific application. Robustness analysis of FDD models against sensor noise under SNR = 3 dB is conducted by injecting controlled noise into the sensor data. The process of this robustness analysis, depicted in Figure 3, involves two assumptions for generating and injecting controlled noise, elaborated as follows:

1) We use Additive White Gaussian Noise (AWGN) to model and generate sensor noise. AWGN is a basic noise model and is chosen because it is a very common source of noise in the real world [26]. The variance of AWGN is consistently equal to its noise power due to its Gaussian random variable nature with zero mean.
2) Noise is only injected into the samples of the testing dataset. Because if noise is added to the samples of their training dataset, the FDD models learn based on these samples and their robustness against sensor noise cannot be calculated anymore.

Algorithm 1 outlines the process of noise generation with SNR = 3 dB and injection into the testing set. *Numpy.random.normal(loc, scale, size)* is used to generate AWGN which is a function provided by the *NumPy* library in *Python*, where *'loc'* represents the mean, *'scale'* represents the standard deviation, and *'size'* determines the number of random samples to generate with the option to specify the output array's dimensions (N). Here, $i$ refers to the $i$-th sample, and N represents the total number of samples in the testing set. This algorithm ensures accurate robustness analysis of FDD models against sensor noise.

---

***Algorithm 1 (Noise generation with SNR = 3 dB and injection into the data of the most important sensor.)***

---

*Input: Original signal (The samples of the unnoisy testing set):* $s_i$

*Output: Noisy signal (The noisy testing set):* $S_i^{'}$

*Generating AWGN that matches the dimensions of the input signal:* $n_i$

1: $P_{signal} = \frac{1}{N}\sum_{i=0}^{N} S_i^2$

2: $P_{noise} = \frac{P_{signal}}{10^{SNR}}$

3: $n_i = np.random.normal(0, \sqrt{P_{noise}}, N)$

*Injecting the generated AWGN into the original signal.*

4: $S_i^{'} = n_i + s_i$

---

## 4. Result and Discussion

### 4.1. Data Preprocessing

Data preprocessing plays a crucial role in influencing the effectiveness of FDD models [27]. To ensure appropriate FDD models building, we conducted data preprocessing encompassing sample labeling, dataset resampling, and splitting. Sample labeling is vital for tree-based ML algorithms, as they are supervised learning techniques necessitating labeled data for FDD model training. Correctly labeling each sample as a specific fault or non-fault class from Table 2 is imperative. Class 0 signifies non-faulty data, whereas other classes are detailed in Table 2.

A class imbalance is apparent in Table 2, which can undermine the performance of FDD models by prioritizing majority classes and neglecting minority ones. To address this, undersampling was applied, decreasing the size of the majority class while preserving all minority data [27]. The final preprocessing step involved splitting the data into 75% training and 25% testing sets, resulting in 415,721 samples for training and 138,574 for testing.

Table 2. The classes of samples and percentage of each class in the dataset.

| Type | Class | Occurrence Percentage |
|---|---|---|
| Non-faulty condition | 0 | 45.6 |
| Open LT display case door | 1 | 9.1 |
| Ice accumulation on LT evaporator coil | 2 | 8.9 |
| LT evaporator expansion valve failure | 3 | 9.1 |
| MT evaporator fan motor failure | 4 | 9.1 |
| Condenser air path blockage | 5 | 9.1 |
| MT evaporator air path blockage | 6 | 9.1 |

### 4.2. Ranking the Sensors based on Feature Importance

After data preprocessing, we constructed four data-driven FDD models using all dataset features by employing tree-based ML algorithms as classifiers, including RF, XGBoost, CatBoost, and LightGBM. These models were built to classify $Co_2$-RS faults, and the feature importance ranking of sensors was extracted through the inherent ranking capabilities of tree-based ML algorithms. Figure 4 presents a comparison of the performance of constructed FDD models based on F1_Score. According to Figure 4, all models perform exceptionally well, with scores above 99%. The FDD model utilizing the LightGBM as a classifier demonstrates the highest performance, achieving an F1_Score of 99.71%, as illustrated in Figure 4. In contrast, the FDD model that employed the CatBoost shows the lowest performance compared to other constructed FDD models. Based on these FDD models, we determined the importance of all sensors in fault classification and ranked them. Figure 5 shows the top-ranked important sensors identified by the constructed FDD models. Detailed results are provided in the supplementary material, in Figure A1 in the appendix. Based on Figure 5, the point is that the condenser-side sensors ($T_{FI}$, $T_{FO}$, $T_C$) are present as the most important sensors in the subset of all classifiers, and this point shows that the occurrence of faults in the system has a significant impact on the condenser.

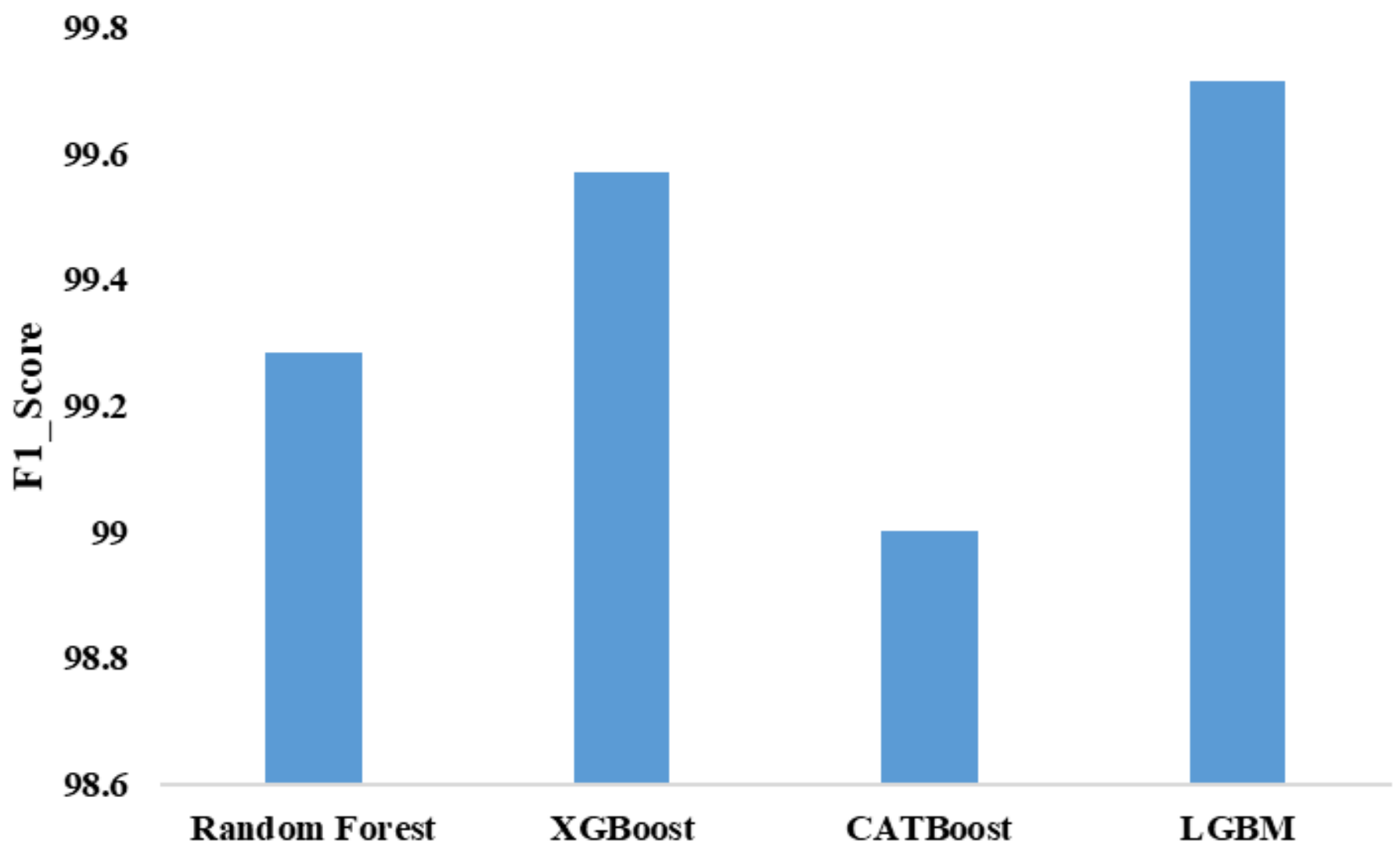

Figure 4. The performance of constructed data-driven FDD models based on F1-score.

### 4.3. Optimal Sensor Set Selection and Robustness Analysis

Using the feature importance rankings derived from the FDD models constructed in Section 4.2, we sorted the sensors in descending order of importance. Sensors were then incrementally added (one by one) to a new dataset based on this ranking. For each iteration, the new dataset was split into training and testing sets. As in the previous section, once again, tree-based ML algorithms— RF, XGBoost, CatBoost, and LightGBM—were trained on the training set to build an FDD model (green line in Figure 3). In the next step, each model's performance was evaluated in two processes: first, using an unnoisy testing set (purple line in Figure 3), and second, using a noisy testing set with AWGN injected at a SNR of 3 dB, as outlined in Section 3.4, to assess robustness (red line in Figure 3). In both cases, the F1-score served as the evaluation metric. For each iteration, noise was injected exclusively into the most important sensor's data within the unnoisy testing set for each classifier. For instance, noise is injected only into condenser inlet air temperature ($T_{FI}$) sensor data in the unnoisy testing set for FDD models which applied RF, CatBoost, and LightGBM as a classifier. While for XGBoost classifier, noise is injected only into the MT evaporator supply air temperature ($T_{Sup1}$). The performance of the FDD model without noise injection was compared against a predefined stopping criterion of 99% F1-score at each sensor addition step. If unmet, the RFA process continued by adding the next ranked sensor, updating the dataset, and repeating until the threshold was achieved. Upon meeting the criterion, the most important and influential sensors are selected to form the optimal sensor set. Figure 3 illustrates all the processes used to select the optimal sensor set by each classifier as a flowchart, while Figure 5 summarizes the performance and robustness results for each classifier.

For example, in the previous section (Section 3.4), RF classifier identified $T_{FI}$ as the most important sensor. In the first RFA iteration, a dataset with only $T_{FI}$ was created, split, and used to train an RF-based FDD model. Evaluated on the unnoisy testing set, it achieved an F1-score of 68.93% (Figure 5a). With noise injected into $T_{FI}$ at SNR = 3 dB, robustness was 14.35%. As the 99% threshold was not met, the RFA process iterated, adding the next sensor (condenser outlet temperature, $T_{FO}$). After four iterations, an optimal set of four sensors was selected, yielding a model with 99% performance and 49.46% robustness. This demonstrates that RF can effectively classify six $CO_2$-RS faults with a minimal sensor set, balancing accuracy and noise resilience.

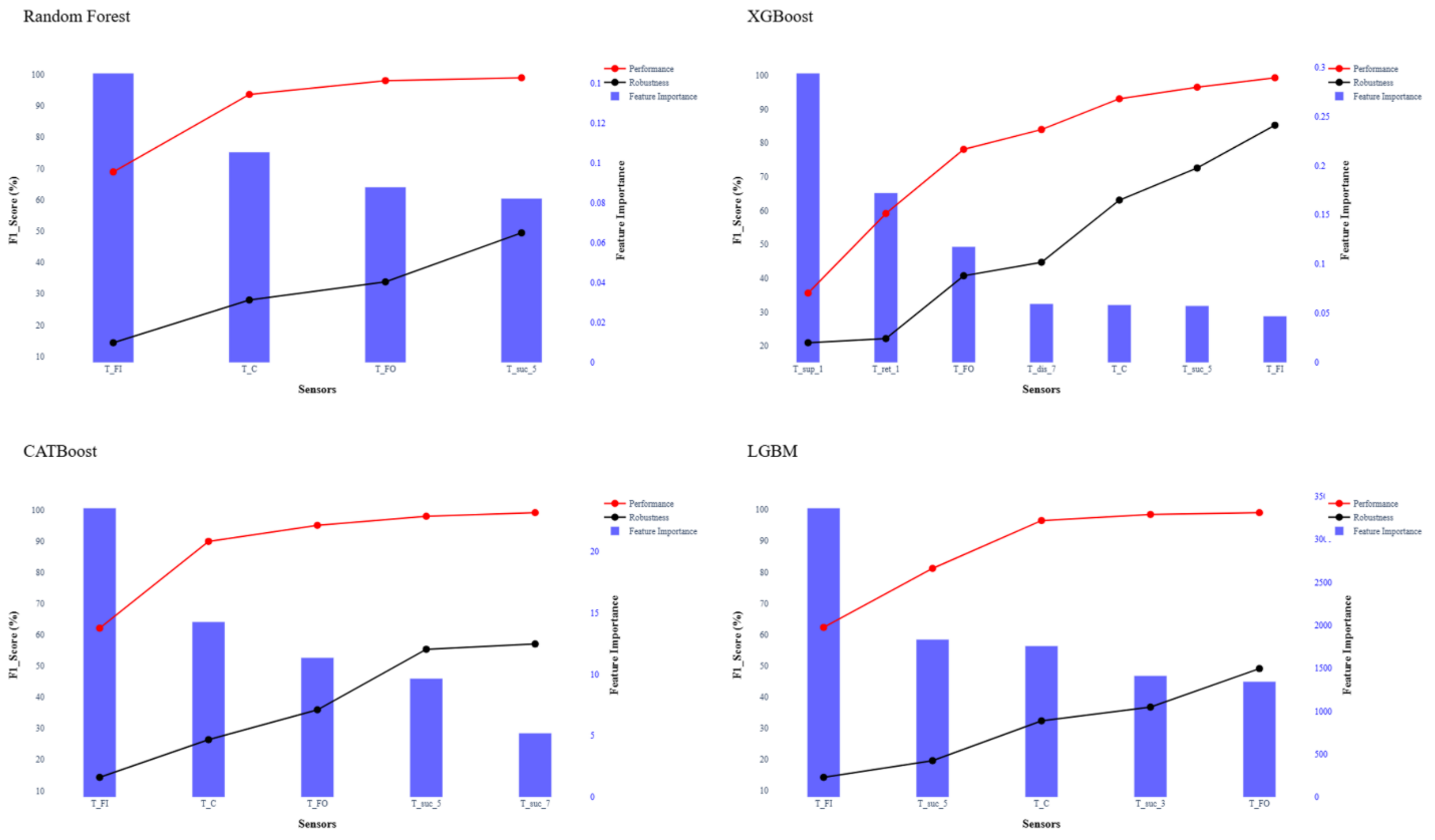

Figure 5. The evaluation results and robustness of the FDD models obtained by the RFA process and tree-based classifiers.

Figure 5 illustrates the optimal sensor sets, performance, and robustness of FDD models constructed using tree-based classifiers (RF, XGBoost, CatBoost, and LightGBM). As the number of sensors increases, the performance of the FDD models improves, reaching the stopping criterion of 99% F1-score. However, for RF and CATBoost classifiers, this increasing trend becomes smoother after the second most important sensor is added. Unlike the XGBoost and LGBM classifiers, the performance of RF and CATBoost improves at a slower rate. Robustness also improves with additional sensors, suggesting that classifiers benefit from diverse data inputs. Even with noise injected into the most important sensor, the presence of non-noisy sensors enables fault classification. Notably, RF achieves a 68.93% F1-score using only the condenser inlet air temperature sensor (T_FI) without noise, but its robustness drops sharply to 14.35% when noise is injected at an SNR of 3 dB. In contrast, XGBoost's initial performance with its most important sensor (T_sup_1) is lower, at approximately 35.6%, with a robustness of 19.54%, indicating greater sensitivity to noise in early iterations compared to other classifiers.

Figure 6 complements this analysis by detailing the number of sensors in the optimal sets, performance, and robustness of FDD models built with tree-based classifiers. RF, XGBoost, CatBoost, and LGBM achieve the 99% performance threshold with optimal sets of 4, 7, 5, and 5 sensors, respectively. RF's smaller set highlights its efficiency in meeting the accuracy target, while XGBoost attains the highest F1-score of 99.25% with seven sensors, reflecting its ability to leverage a larger configuration for optimal results. This reduced dependence on any single sensor enhances XGBoost's robustness, peaking at 85.24%, followed by CatBoost at 57.07%. The gradient-boosting framework of XGBoost effectively captures complex, non-linear data relationships, contributing to its stability under noisy conditions. This superior performance and robustness position XGBoost as a reliable choice for CO2-RS FDD. Moreover, the use of a modest number of sensors, predominantly cost-effective temperature sensors, improves system efficiency and reduces initial and maintenance costs, offering a practical solution for refrigeration system monitoring.

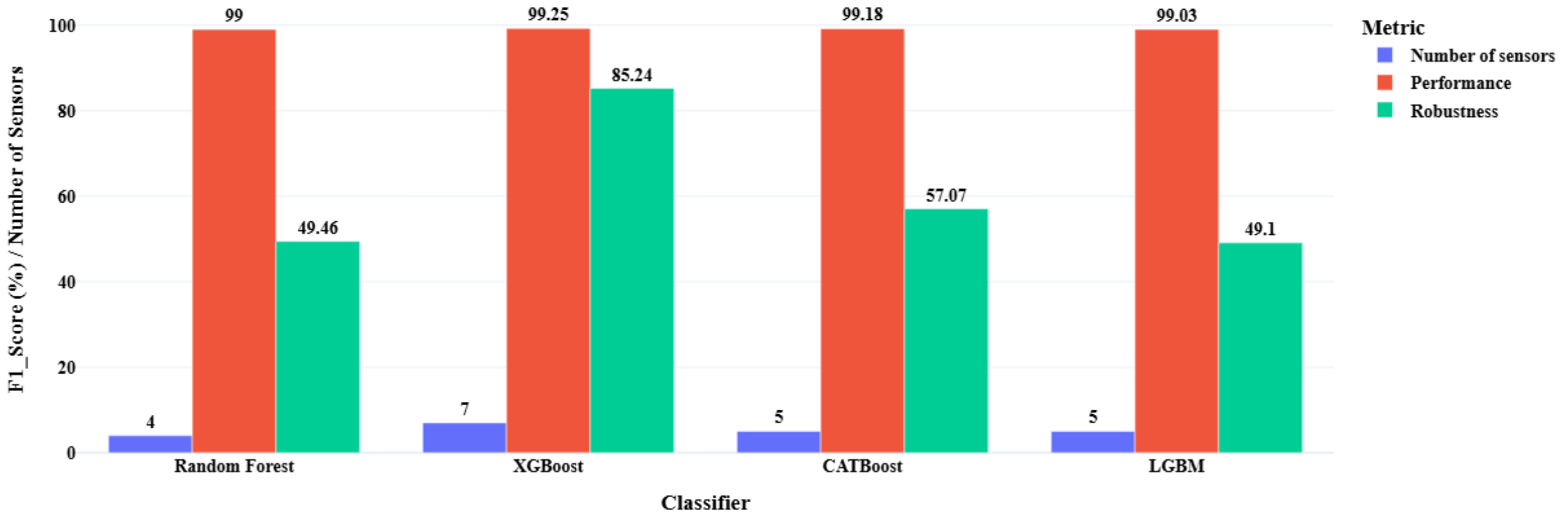

Figure 6. Performance and robustness of optimal sensor sets across tree-based classifiers.

### 4.4. Sensitivity Analysis of FDD Models Under Different Scenarios

While the primary analysis in this study evaluated the robustness of FDD models and their optimal sensor sets at an SNR of 3 dB, real-world conditions may present a broader range of noise levels and sensor reliability challenges. To further examine the sensitivity and practical applicability of these models, this section investigates the robustness of FDD models, which were constructed using their selected optimal sensor sets in the previous section, under three distinct scenarios: (1) a high-SNR condition (SNR = 10 dB), (2) a low-SNR worst-case condition (SNR = 0 dB), and (3) a complete failure of the most important sensor in the optimal sensor set. The results of these scenarios provide a deeper insight into the FDD models' ability to maintain performance under varying noise conditions and sensor reliability constraints. In the following, we will detail the analysis of each of these three scenarios. Figure 7 illustrates the robustness of FDD models which were constructed using their selected optimal sensor sets in the previous section across scenarios 1 and 2 (SNR = 0 dB, 10 dB, and baseline 3 dB).

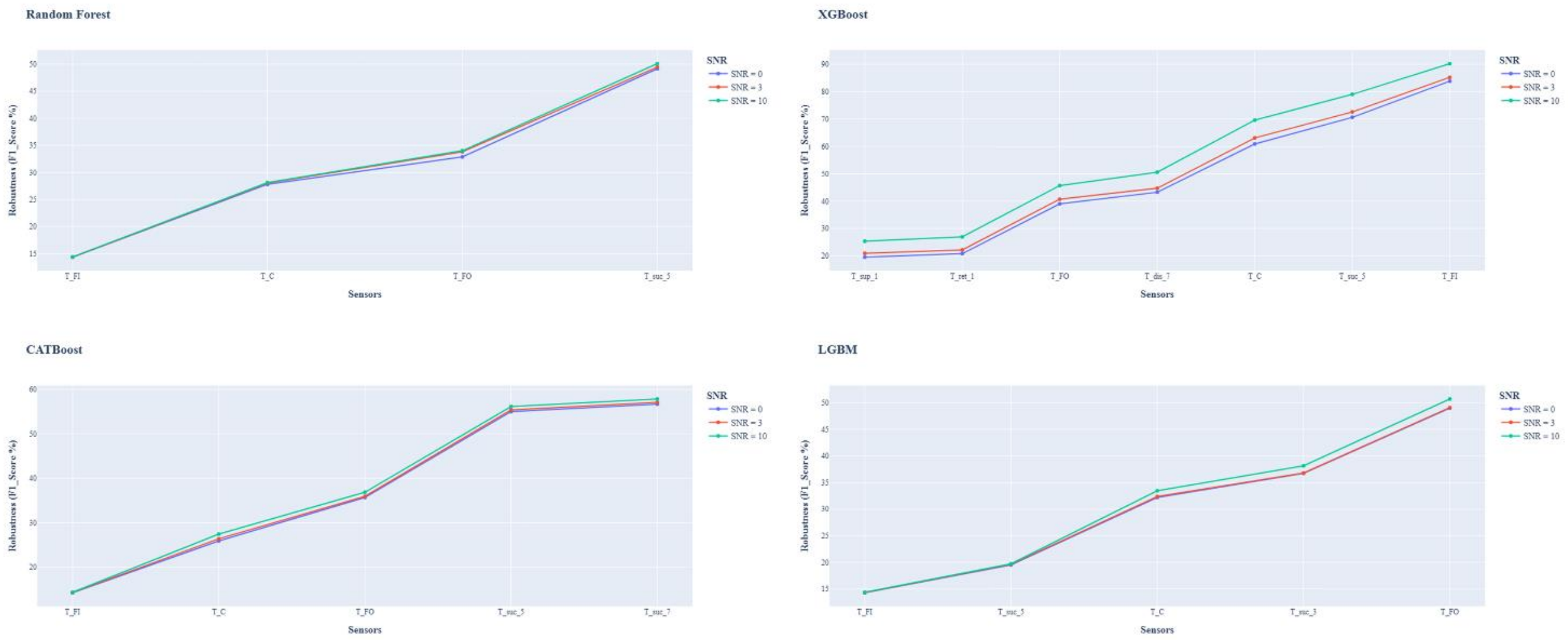

Figure 7. Robustness of FDD models and optimal sensor sets across scenarios 1 and 2 (SNR = 0 dB, 10 dB, and baseline 3 dB).

### 4.4.1. Scenario 1: Robustness at Higher-SNR Condition (SNR = 10 dB)

In the first scenario, the robustness of the FDD models was assessed at an SNR of 10 dB, where the signal power is ten times the noise power, representing a high-quality signal environment. Across all classifiers— RF, XGBoost, CatBoost, and LightGBM—the optimal sensor sets demonstrated improved robustness scores compared to the baseline SNR of 3 dB, as shown in Figure 7. For RF, as shown in Figure 7a, the robustness score increased marginally from 33.82% with three sensors ($T_{FI}$, $T_C$, $T_{FO}$) at SNR = 3 dB to 34.01% at SNR = 10 dB, reflecting a modest gain due to its smaller sensor set. XGBoost, with its seven-sensor configuration ($T_{Sup1}$, $T_{ret1}$, $T_{FO}$, $T_{dis7}$, $T_C$, $T_{Suc5}$ , $T_{FI}$), exhibited a more pronounced improvement, rising from 85.24% at SNR = 3 dB to 90.23% at SNR = 10 dB, as shown in Figure 7b. In other words, the most significant improvement was observed in XGBoost, where $T_{FI}$ improved from 85.24% to 90.23%, demonstrating that reducing noise interference enhances model performance. Similarly, CatBoost and LightGBM showed gains, with robustness scores reaching 57.84% (five sensors) and 50.71% (five sensors), respectively, at SNR = 10 dB. Figure 7c and 7d confirm this. These results indicate that higher signal quality enhances the models' ability to maintain diagnostic accuracy, with XGBoost benefiting most due to its larger sensor set and gradient-boosting framework, which excels at leveraging cleaner data.

#### 4.4.2. Scenario 2: Robustness at Lower-SNR or Worst-Case Condition (SNR = 0 dB)

In contrast to Scenario 1, this scenario investigates the worst-case condition where the SNR is reduced to 0 dB, meaning the signal power is equal to the noise power. This condition significantly degrades the sensor data quality, making fault detection more challenging. As shown in Figure 7, the robustness of all FDD models declined significantly compared to the SNR = 3 dB baseline, highlighting the challenges of operating in highly noisy environments. RF's robustness dropped to 32.87% with three sensors, a decrease from 33.82%, underscoring its sensitivity to noise with a minimal sensor set. XGBoost, despite the noise, maintained a robustness score of 83.83% with seven sensors, down from 85.24%, demonstrating superior resilience attributable to its iterative error correction and larger sensor ensemble. CatBoost and LightGBM recorded robustness scores of 56.68% (five sensors) and 48.99% (five sensors), respectively, at SNR = 0 dB, reflecting moderate declines from their SNR = 3 dB performances (56.68% and 49.1%). These findings emphasize that while all models suffer in low-SNR conditions, XGBoost's robustness makes it the most reliable choice for fault detection in noisy settings, though even it struggles as noise overwhelms the signal.

#### 4.4.3. Scenario 3: Complete Failure of the Most Important Sensor

The third scenario simulates a complete failure of the most important sensor in each optimal set, a critical test of the models' reliance on key sensors and their overall reliability. A complete failure was modeled by zeroing the data of the top-ranked sensor (e.g., $T_{FI}$ for RF, CatBoost, and LightGBM; $T_{Sup1}$ for XGBoost) in the testing set, mimicking a hardware malfunction where the sensor outputs a constant, uncorrelated value. The results indicate a severe impact on FDD model robustness, as shown in Figure 8. Without failure, the baseline F1-scores were 99% (RF), 99.25% (XGBoost), 99.18% (CatBoost), and 99.03% (LightGBM). Upon failure, robustness plummeted across all models, with RF declining to 37% (a 62% drop), CatBoost to 44% (55.18% drop), and LightGBM to 41% (58.03% drop). XGBoost, however, retained a robustness score of 79%, with a relatively modest decline of 20.25%, highlighting its reduced dependence on any single sensor due to its broader optimal set and adaptive boosting mechanism. This stark contrast underscores that while minimal sensor sets (e.g., RF's three sensors) achieve high performance under ideal conditions, they are highly vulnerable to the loss of a critical sensor. XGBoost's resilience suggests that a slightly larger, well-distributed sensor set enhances reliability, offering a trade-off between cost and robustness in real-world applications where sensor failures are plausible.

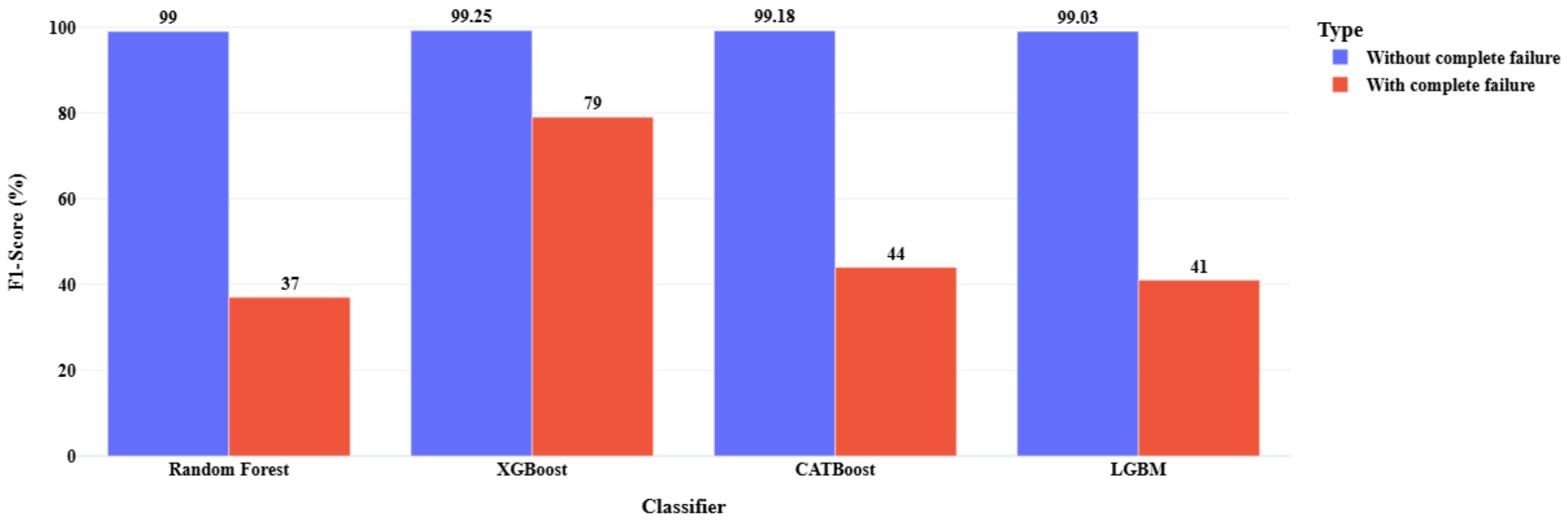

Figure 8. Impact of critical sensor failure on FDD model robustness.

The sensitivity analysis reveals that FDD model performance is intricately tied to both signal quality and sensor reliability. At SNR = 10 dB, all models benefit from cleaner data, with XGBoost achieving the highest robustness, reinforcing its suitability for environments with controlled noise levels. Conversely, the SNR = 0 dB scenario exposes vulnerabilities, particularly for models with fewer sensors like RF, while XGBoost's sustained performance highlights the value of a robust sensor ensemble in adverse conditions. The complete failure scenario further amplifies this disparity, demonstrating that minimal sensor sets, though cost-effective, risk catastrophic performance drops when a key sensor fails—a critical consideration for supermarket $CO_2$-RS deployments where downtime and maintenance costs are significant. These insights suggest that while the baseline SNR = 3 dB analysis provides a balanced evaluation, real-world FDD implementations must account for variable noise levels and potential sensor failures. Future research could explore adaptive SNR thresholds or redundant sensor configurations to mitigate these risks, ensuring sustained performance across diverse operational contexts.

## Conclusion

This study advances the field of Fault Detection and Diagnosis (FDD) in $CO_2$ refrigeration systems ($CO_2$-RS) by developing robust, data-driven models using tree-based machine learning algorithms—Random Forest (RF), XGBoost, CatBoost, and LightGBM—to classify six prevalent faults in a laboratory-scale system. Our primary findings underscore the efficacy of the Recursive Feature Addition (RFA) approach in identifying optimal sensor sets, achieving a 99% F1-score threshold with minimal sensors: four for RF, seven for XGBoost, five for CatBoost, and five for LightGBM. This optimization reduces implementation costs while maintaining high diagnostic accuracy, addressing a critical research gap in sensor selection for FDD models. Notably, condenser-side sensors (e.g., $T_{FI}$, $T_C$, $T_{FO}$) consistently ranked among the most influential, highlighting their pivotal role in fault detection across all classifiers.

Robustness analysis against sensor noise, conducted by injecting Additive White Gaussian Noise (AWGN) at a signal-to-noise ratio (SNR) of 3 dB into the most critical sensor, revealed XGBoost's superior resilience, retaining 85.24% robustness with its seven-sensor ensemble. Sensitivity analysis across high-SNR (10 dB), low-SNR (0 dB), and sensor failure scenarios further validated XGBoost's stability, peaking at 90.23% robustness at 10 dB and retaining 79% under critical sensor failure, compared to sharper declines in RF (37%), CatBoost (44%), and LightGBM (41%). These results illuminate a trade-off between sensor count, cost, and reliability: minimal sets enhance efficiency but increase vulnerability, while larger ensembles bolster noise resilience and fault tolerance.

The study bridges key research gaps by integrating optimal sensor selection with comprehensive robustness analysis, offering a practical framework for enhancing energy efficiency and fault management in supermarket $CO_2$-RS. Future research could explore adaptive SNR thresholds tailored to specific operational contexts, mitigating performance variability under diverse noise conditions. Additionally, incorporating redundant sensor configurations or real-time recalibration strategies could further enhance reliability against sensor failures. Extending this methodology to full-scale commercial systems and diverse fault types would validate its scalability, while investigating hybrid ML approaches might refine diagnostic precision. These advancements promise to bolster FDD adoption, reducing energy waste and greenhouse gas emissions in the supermarket sector.

## Conflicts of interest

There are no conflicts of interest to declare.

## Data availability

Data will be made available on request.

## Acknowledgment

The research work was not supported by any foundation.

## Author Contributions

**Masoud Kishani Farahani**: Conceptualization; Investigation; Methodology; Validation; Visualization; Writing - Original Draft. **Morteza Kolivandi**: Visualization; Validation; Writing -

Review & Editing. **Abbas Rajabi Ghahnavieh**: Conceptualization; Supervision; Validation; Writing - Review & Editing. **Mohammad Talaei**: Validation; Writing - Review & Editing.

## Supplementary Material

Table A1: The list of the installed sensors on the $CO_2$-RS.

| **Symbols** | **Description** | **SI unit** |
|---|---|---|
| $W_1$ | MT 1st compressor power | W |
| $W_2$ | MT 2nd compressor power | W |
| $W_3$ | MT 3rd compressor power | W |
| $W_4$ | LT 1th compressor power | W |
| $W_5$ | LT 2nd compressor power | W |
| $W_6$ | Condenser fan power | W |
| $M_1$ | Flash tank bypass mass flow rate | kg/min |
| $M_2$ | LT evaporator mass flow rate | kg/min |
| $M_3$ | MT evaporator mass flow rate | kg/min |
| $P_{dis1}$ | MT compressor rack outlet pressure | Mpa |
| $P_{suc1}$ | MT compressor rack inlet pressure | Mpa |
| $P_{dis2}$ | LT compressor rack outlet pressure | Mpa |
| $P_{suc2}$ | LT compressor rack inlet pressure | Mpa |
| $P_{dis3}$ | Flash tank vapor outlet pressure | Mpa |
| $P_{suc3}$ | LT display case suction pressure | Mpa |
| $P_{suc4}$ | MT display case suction pressure | Mpa |
| $T_{dis1}$ | MT 1st compressor discharge temperature | °C |
| $T_{suc1}$ | MT 1st compressor suction temperature | °C |
| $T_{dis2}$ | MT 2nd compressor discharge temperature | °C |
| $T_{suc2}$ | MT 2nd compressor suction temperature | °C |
| $T_{dis3}$ | MT 3rd compressor discharge temperature | °C |
| $T_{suc3}$ | MT 3rd compressor suction temperature | °C |
| $T_{dis4}$ | LT 1st compressor discharge temperature | °C |
| $T_{suc4}$ | LT 1st compressor suction temperature | °C |
| $T_{dis5}$ | LT 2nd compressor discharge temperature | °C |
| $T_{suc5}$ | LT 2nd compressor suction temperature | °C |
| $T_{dis6}$ | MT compressor rack outlet temperature | °C |
| $T_{suc6}$ | MT compressor rack inlet temperature | °C |
| $T_{dis7}$ | LT compressor rack outlet temperature | °C |
| $T_{suc7}$ | LT compressor rack inlet temperature | °C |
| $T_{suc8}$ | Flash tank vapor outlet temperature | °C |
| $T_{suc9}$ | LT display case suction temperature | °C |
| $T_{suc10}$ | MT display case suction temperature | °C |
| $T_C$ | Condenser outlet temperature | °C |
| $T_{FI}$ | Condenser inlet air temperature | °C |
| $T_{FO}$ | Condenser outlet air temperature | °C |
| $T_{sup1}$ | MT evaporator supply air temperature | °C |
| $T_{ret1}$ | MT evaporator return air temperature | °C |
| $T_{sup2}$ | LT evaporator supply air temperature | °C |
| $T_{ret2}$ | LT evaporator return air temperature | °C |

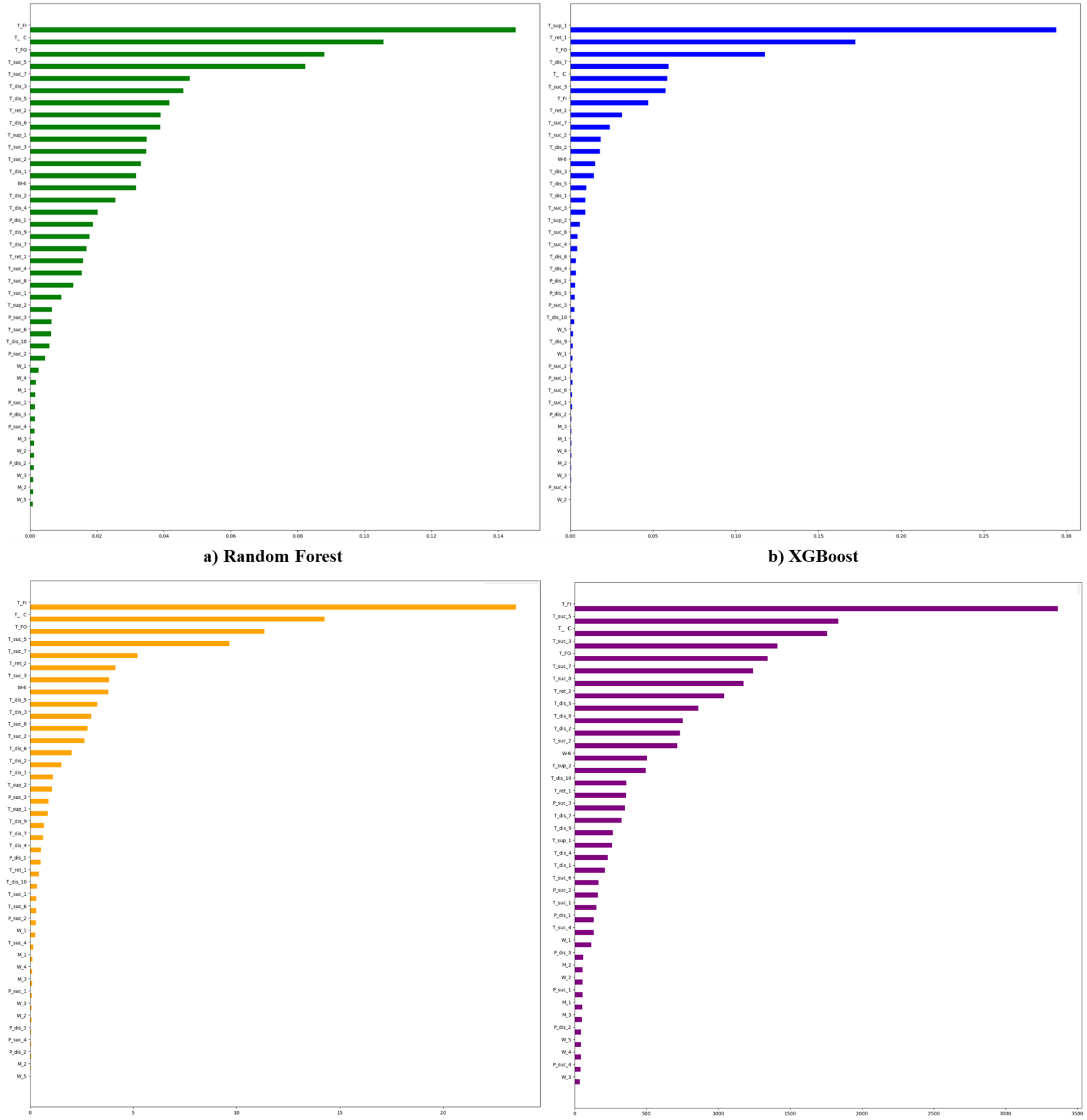

a) Random Forest b) XGBoost c) CATBoost d) LGBM

Figure A1. Feature importance ranking by FDD models based on tree-based classifiers.